\documentclass[aip,pof]{revtex4}
\usepackage{graphicx}
\usepackage{amsmath}
\usepackage{amssymb}
\usepackage{bm}
\graphicspath{{fig/}}
\usepackage{color}

\begin{document}
\title{Quantitative test of the time dependent Gintzburg-Landau
equation for sheared granular flow in two dimension}
\author{Kuniyasu Saitoh}
\email[]{k.saitoh@utwente.nl}
\homepage[]{http://www2.msm.ctw.utwente.nl/saitohk/}
\affiliation{Faculty of Engineering Technology, University of Twente, Enschede, the Netherlands}
\author{Hisao Hayakawa}
\email[]{hisao@yukawa.kyoto-u.ac.jp}
\affiliation{Yukawa Institute for Theoretical Physics, Kyoto University, Sakyo-ku, Kyoto, Japan}
\date{\today}
\begin{abstract}
We examine the validity of the time-dependent Ginzburg-Landau equation of granular fluids
for a plane shear flow under the Lees-Edwards boundary condition derived from a weakly
nonlinear analysis through the comparison with the result of discrete element method.
We verify quantitative agreements in the time evolutions of the area fraction and
the velocity fields, and also find qualitative agreement in the granular temperature.
\end{abstract}
\pacs{45.70.Mg, 45.70.Qj, 47.50.Gj}
\maketitle
\section{Introduction}\label{intro}
Flows of granular particles have been extensively studied due to the importance in powder
technology, civil engineering, mechanical engineering, geophysics, astrophysics, applied
mathematics and physics \cite{luding1,luding2,bri,gold}. The characteristic properties of
the granular flows are mainly caused by the inelastic collisions \cite{jeager}. In particular,
the study of granular gases under a plane shear plays an important role in the application
of the kinetic theory \cite{sela,santos,lun,dufty1,dufty2,lutsko1,lutsko2,lutsko3,jr1,jr2},
the shear band or the plug in a moderate dense flow \cite{tan,saitoh}, the long-time tail and the long-range
correlations \cite{kumaran1,kumaran2,kumaran3,orpe1,orpe2,rycroft,lutsko0,otsuki0,otsuki1,otsuki2},
the pattern formation of dense flow \cite{louge1,louge2,louge3,louge4,khain1,khain2},
the determination of the constitutive equation for dense flow \cite{midi,cruz,hatano1},
as well as jamming transition \cite{hecke,hatano2,hatano3,otsuki3,otsuki4,otsuki5,otsuki6}.

The granular hydrodynamic equations based on the kinetic theory well describe the
dynamics of moderate dense granular gases \cite{lun,dufty1,dufty2,lutsko1,lutsko2,lutsko3,jr1,jr2},
even though its applicability is questionable because of the lack of scale separation and the
existence of long range correlations, etc. The two-dimensional granular shear flow is an
appropriate target to check the validity of the granular hydrodynamic equations, where
two denser regions are formed near the boundaries and collide to form a single dense
plug under a physical boundary condition \cite{tan,saitoh}. We refer the dense plugs
as \emph{shear bands} throughout this paper (even though "shear-band" is often referred
to the region of lower density with higher shear rate in the literature of engineering).
A similar shear band is also observed under
the Lees-Edwards boundary condition. The transient dynamics of the shear band and the
hydrodynamic fields can be described by the granular hydrodynamic equations, where reasonable
agreements with the discrete element method (DEM) simulation have been verified \cite{saitoh}. 
It is also known that a homogeneous state of the two-dimensional granular shear flow is
intrinsically unstable as predicted by the linear stability analysis
\cite{linear1,linear2,layer1,layer2,layer3,layer4,layer5}.

To understand the shear band formation after the homogeneous state becomes unstable,
we have to develop the weakly nonlinear analysis. Recently, Shukla and Alam carried
out a weakly nonlinear analysis of the sheared granular flow in finite size systems,
where they derived the Stuart-Landau equation for the disturbance amplitude of the
hydrodynamic fields under a physical boundary condition \cite{shukla1,shukla2,shukla3,add-alam1,add-alam2}.
They found the existence of subcritical bifurcation in both dilute and dense regimes,
while a supercritical bifurcation appears in the medium regime and the extremely
dilute regime. The Stuart-Landau equation, however, does not include any spatial degrees
of freedom and cannot be used to study the slow evolution of the spatial structure of
shear band. We also notice that the shear rate is fixed to unity and cannot be used as a
control parameter in their analysis.

It is also notable that several authors found coexistence of solid and liquid
phases in their molecular dynamics simulations of dense granular shear flows
\cite{khain1,khain2,add-alam3,add-alam4,add-alam5,add-alam6}. In particular, Khain
showed a hysteresis loop of the order parameter defined as a density contrast
between the boundary and the center region \cite{khain1,khain2}. It should be
noted, however, that the mechanism of the subcritical bifurcation based on a
set of hydrodynamic equations differs from that observed in the jamming transition
of frictional particles \cite{otsuki6}.
Indeed, the hysteresis loop in the jamming, which is observed for polydisperse grains,
is originated from the frustrated and metastable configurations of frictional grains,
while the hysteresis for monodisperse grains observed by Khain is from the coexistence
of a crystal structure and a liquid structure.

In our previous work, we have developed the weakly nonlinear analysis for the two-dimensional
granular shear flow and derived the time dependent Ginzburg-Landau (TDGL) equation for
the disturbance amplitude. We introduced a hybrid approach to the weakly nonlinear analysis,
where the derived TDGL equation is written as a two-dimensional form and has time dependent
diffusion coefficients \cite{saitoh1}. We have also discussed the bifurcation of the amplitude,
however, the studies of the numerical solution of the TDGL equation and comparison with the
DEM simulation had been left as an incomplete part of our previous paper \cite{saitoh1}.
Part of this study without comparison with DEM simulation has been published in another paper \cite{RGD28}

In this paper, we quantitatively examine the validity of the derived TDGL equation
for a two dimensional granular shear flow from the comparison with the DEM simulation.
In Sec. \ref{sec:overview}, we review the weakly nonlinear analysis and the hybrid
approach. In Sec. \ref{sec:dem}, which is the main part of this paper, we compare
the numerical solutions of the TDGL equation with the results of DEM simulation.
In Sec. \ref{sec:summ}, we discuss and conclude our results.

\section{Overview of weakly nonlinear analysis}\label{sec:overview}

In this section, we review our previous results for the weakly nonlinear analysis, where
the time evolution for the disturbance amplitude is described by the TDGL equation \cite{saitoh1}. 
We also apply the hybrid approach to the TDGL equation to describe the structural changes
of the shear band \cite{saitoh1}. In Sec. \ref{sub:basic}, we introduce the basic equations.
In Sec. \ref{sub:weakly}, we review the weakly nonlinear analysis to derive the TDGL equation.
In Sec. \ref{sub:hybrid}, we derive a two-dimensional TDGL equation adopting the hybrid
approach to the weakly nonlinear analysis.

\subsection{Basic equations}\label{sub:basic}

Let us explain our setup and basic equations. To avoid difficulties caused by the physical
boundary condition, we adopt the Lees-Edwards boundary condition \cite{lees}, where the upper
and the lower image cells move to the opposite directions with a constant speed $U/2$.
Here, the distance between the upper and the lower image cells is given by $L$. 
We assume that the granular disks are identical, where the mass, the diameter and the restitution coefficient
are respectively given by $m$, $d$ and $e$. In the following argument, we scale the mass,
the length and the time by $m$, $d$ and $2d/U$, respectively. Therefore, the shear rate $U/L$
is nondimensionalized as $\epsilon \equiv 2d/L$ which becomes a small parameter in the
hydrodynamic limit $L\gg d$.

We employ a set of hydrodynamic equations of granular disks derived by Jenkins and
Richman \cite{jr1}. Although their original equations include the angular momentum and the
spin temperature, it is known that the spin effects are localized near the boundary \cite{mitarai}
and the effect of rotation can be absorbed in the normal restitution coefficient,
if the friction constant is small \cite{jz,yj,saitoh}. Thus, our system is reduced to a system
without the spin effects and the dimensionless hydrodynamic equations are given by
\begin{eqnarray}
\left( \partial_t + \mathbf{v} \cdot \nabla \right)
\nu &=& - \nu \nabla \cdot \mathbf{v} \label{eq:hy1} \\
\nu \left( \partial_t + \mathbf{v} \cdot \nabla \right)
\mathbf{v} &=& - \nabla \cdot \mathsf{P} \label{eq:hy2} \\
\left( \nu / 2 \right) \left( \partial_t + \mathbf{v} \cdot \nabla \right)
\theta &=& - \mathsf{P} : \nabla \mathbf{v} - \nabla \cdot \mathbf{q} - \chi \label{eq:hy4}~,
\end{eqnarray}
where $\nu$, $\mathbf{v}=(u,w)$, $\theta$, $t$ and $\nabla = (\partial / \partial_x, \partial / \partial_y)$
are the area fraction, the dimensionless velocity fields, the dimensionless granular
temperature, the dimensionless time and the dimensionless gradient, respectively.
The pressure tensor $\mathsf{P}=(P_{ij})$, the heat flux  $\mathbf{q}$ and the energy
dissipation rate $\chi$ are given in the dimensionless forms as
\begin{eqnarray}
P_{ij} &=& \left[ p(\nu)\theta - \xi(\nu)\theta^{1/2}
\left( \nabla \cdot \mathbf{v} \right) \right] \delta_{ij} - \eta(\nu)\theta^{1/2} e_{ij}~,
\label{eq:stress} \\
\mathbf{q} &=& - \kappa(\nu) \theta^{1/2} \nabla \theta - \lambda(\nu) \theta^{3/2} \nabla \nu~,
\label{eq:heat} \\
\chi &=& \frac{1-e^2}{4\sqrt{2\pi}}\nu^2 g(\nu) \theta^{1/2} \left[ 4\theta - 3\sqrt{\frac{\pi}{2}}
\theta^{1/2} \left( \nabla \cdot \mathbf{v} \right) \right]~,
\label{eq:chi}
\end{eqnarray}
respectively, where $p(\nu)\theta$, $\xi(\nu)\theta^{1/2}$, $\eta(\nu)\theta^{1/2}$, $\kappa(\nu)\theta^{1/2}$
and $\lambda(\nu)\theta^{3/2}$ are the dimensionless forms of the static pressure, the bulk
viscosity, the shear viscosity, the heat conductivity and the coefficient associated with the
gradient of density, respectively, and $e_{ij}\equiv(\nabla_jv_i+\nabla_iv_j-\delta_{ij}\nabla\cdot\mathbf{v})/2$
$(i,j = x,y)$ is the deviatoric part of the strain rate tensor. 
The explicit forms of them are listed
in Table \ref{tab:subfunctions}, where we adopt the radial distribution function at contact
\begin{equation}
g(\nu) = \frac{1-7\nu/16}{\left( 1-\nu \right)^2}~,
\label{eq:radial}
\end{equation}
which is only valid for $\nu < 0.7$ \cite{gnu4,gnu3,gnu2,gnu1}.
\begin{table}
\begin{tabular}{lll}
\hline
$p(\nu)$ &$=$& $\frac{1}{2}\nu \left[ 1+(1+e)\nu g(\nu) \right]$ \\
$\xi(\nu)$ &$=$& $\frac{1}{\sqrt{2\pi}} (1+e)\nu^2 g(\nu)$ \\
$\eta(\nu)$ &$=$& $\sqrt{\frac{\pi}{2}} \left[ \frac{g(\nu)^{-1}}{7-3e}
+ \frac{(1+e)(3e+1)}{4(7-3e)}\nu + \left( \frac{(1+e)(3e-1)}{8(7-3e)} + \frac{1}{\pi} \right)(1+e)\nu^2 g(\nu) \right]$ \\
$\kappa(\nu)$ &$=$& $\sqrt{2\pi} \left[ \frac{g(\nu)^{-1}}{(1+e)(19-15e)}
+ \frac{3(2e^2+e+1)}{8(19-15e)}\nu + \left( \frac{9(1+e)(2e-1)}{32(19-15e)} + \frac{1}{4\pi} \right)(1+e)\nu^2 g(\nu) \right]$ \\
$\lambda(\nu)$ &$=$& $- \sqrt{\frac{\pi}{2}}\frac{3e(1-e)}{16(19-15e)}\left[ 4(\nu g(\nu))^{-1} + 3(1+e) \right] \frac{d \left( \nu^2g(\nu) \right)}{d\nu}$ \\
\hline
\end{tabular}
\caption{The functions in Eqs.(\ref{eq:stress})-(\ref{eq:chi}).}
\label{tab:subfunctions}
\end{table}
\subsection{Weakly nonlinear analysis}\label{sub:weakly}

To study the slow dynamics of shear band, we need to develop a weakly nonlinear analysis.
For this purpose, we introduce a long time scale $\tau\equiv\epsilon^2 t$ and long length
scales $(\xi,\zeta)\equiv\epsilon (x,y)$. We also introduce the neutral solution around
the most unstable mode $\mathbf{q}_c=(0,q_c)$ as
\begin{equation}
\hat{\phi}_\mathrm{n} = A^{\rm L}(\zeta,\tau)\phi^{\rm L}_{q_c}e^{i q_c \zeta}
+ \mathrm{c.c.}~, \label{eq:neutral}
\end{equation}
where $\mathrm{c.c.}$ represents the complex conjugate and $\phi^{\rm L}_{q_c}$
corresponds to the Fourier coefficient of the hydrodynamic fields at $\mathbf{q}_c$. We notice that
the amplitude of the layering
mode $A^{\rm L}(\zeta,\tau)$ depends on $\zeta$ but is independent of $\xi$, because
any non-layering modes $q_x\neq 0$ are linearly stable. 
Then, we expand $A^{\rm L}(\zeta,\tau)$
into the series of $\epsilon$ as
\begin{equation}
A^{\rm L}(\zeta,\tau) = \epsilon A^{\rm L}_1 + \epsilon^2 A^{\rm L}_2
+ \epsilon^3 A^{\rm L}_3 + \dots~. \label{eq:amp_exp}
\end{equation}
Substituting Eqs. (\ref{eq:neutral}) and (\ref{eq:amp_exp}) into the hydrodynamic
equations (\ref{eq:hy1})-(\ref{eq:hy4}) and collecting terms in each order of $\epsilon$,
we obtain an amplitude equation.

The first non-trivial equation at $O(\epsilon^3)$ is the TDGL equation
\begin{equation}
\partial_\tau A^{\rm L}_1 = \sigma_c A^{\rm L}_1 + D \partial_\zeta^2 A^{\rm L}_1
+ \beta A^{\rm L}_1 |A^{\rm L}_1|^2 , \label{eq:GL30}
\end{equation}
where $D$ and $\beta$ are listed in Table 2 of Ref. \cite{saitoh1}. Here,
$\sigma_c$ is the maximum growth rate at $\mathbf{q}_c$ scaled by $\epsilon^2$. 
Because
of the scaling relations $D=\bar{D}$ and $\beta=\epsilon\bar{\beta}$, we can rewrite the
TDGL equation as the equation for the scaled amplitude $\bar{A}^{\rm L}_1\equiv\epsilon^{1/2}A^{\rm L}_1$ as
\begin{equation}
\partial_{\tau} \bar{A}^{\rm L}_1 = \sigma_c \bar{A}^{\rm L}_1
+ \bar{D} \partial_{\zeta}^2 \bar{A}^{\rm L}_1 +
\bar{\beta} \bar{A}^{\rm L}_1 |\bar{A}^{\rm L}_1|^2~.
\label{eq:red_TDGL3_layer}
\end{equation}
It should be noted that the TDGL equation (\ref{eq:GL30}) or (\ref{eq:red_TDGL3_layer})
can be only used for $\beta, \bar{\beta}<0$, i.e., the case of a supercritical bifurcation.

Developing a similar procedure till $O(\epsilon^5)$, we also obtain the amplitude equation
\begin{equation}
\partial_\tau \check{A}^{\rm L} =
\sigma_c \check{A}^{\rm L} + \bar{D} \partial_\zeta^2 \check{A}^{\rm L}
+ \bar{\beta} \check{A}^{\rm L} |\check{A}^{\rm L}|^2
+ \epsilon \bar{\gamma} \check{A}^{\rm L} |\check{A}^{\rm L}|^4 + O(\epsilon^3)~,
\label{eq:GL5_red}
\end{equation}
where we have introduced $\check{A}^{\rm L}(\zeta,\tau) = \epsilon^{1/2} [A^{\rm L}_1(\zeta,\tau) +
\epsilon A^{\rm L}_2(\zeta,\tau) + \epsilon^2 A^{\rm L}_3(\zeta,\tau)]$ and
$\bar{\gamma}$ is also listed in Table 2 of Ref. \cite{saitoh1}. Equation (\ref{eq:GL5_red})
can be used for $\bar{\beta}>0$ and $\bar{\gamma}<0$, i.e., the case of a subcritical bifurcation.

\subsection{Hybrid approach to the weakly nonlinear analysis}\label{sub:hybrid}

Although we derived the TDGL equations (\ref{eq:red_TDGL3_layer}) and (\ref{eq:GL5_red}),
these equations do not include $\xi$ and they are still not appropriate to study the
two-dimensional structure of shear band. Therefore, we need a new approach, where the
non-layering mode is coupled with the layering mode. For this purpose, we add a small
deviation to the most unstable mode as $\mathbf{q}(\tau)=\mathbf{q}_c+\delta\mathbf{q}(\tau)$
and assume $\hat{\phi}_\mathrm{n}$ does not change if the deviation $\delta\mathbf{q}(\tau)$
is small. Then, Eq. (\ref{eq:neutral}) can be rewritten as
\begin{equation}
\hat{\phi}_\mathrm{n} \simeq A^{\rm L}(\xi,\zeta,\tau)\phi^{\rm L}_{q_c}
e^{i \mathbf{q}(\tau)\cdot\mathbf{z}} + \mathrm{c.c.}~, \label{eq:neutral-d}
\end{equation}
where we have introduced $\mathbf{z}\equiv(\xi,\zeta)$ and a $\xi$-dependent amplitude
$A^{\rm L}(\xi,\zeta,\tau)$. If we also take into account the contribution from the
non-layering mode, a hybrid solution is given by
\begin{eqnarray}
\hat{\phi}_\mathrm{h} &=& \left\{ A^{\rm L}(\xi,\zeta,\tau)\phi^{\rm L}_{q_c}
+ A^{\rm NL}(\xi,\zeta,\tau) \phi^{\rm NL}_{\mathbf{q}(\tau)} \right\}
e^{i \mathbf{q}(\tau) \cdot \mathbf{z}} + \mathrm{c.c.} \nonumber\\
&\simeq & A(\xi,\zeta,\tau)\left\{ \phi^{\rm L}_{q_c} +
\phi^{\rm NL}_{\mathbf{q}(\tau)} \right\}e^{i \mathbf{q}(\tau) \cdot \mathbf{z}}
+ \mathrm{c.c.}~, \label{eq:neutral2}
\end{eqnarray}
where $A^{\rm NL}(\xi,\zeta,\tau) $ and $\phi^{\rm NL}_{\mathbf{q}(\tau)}$ are the amplitude
and the Fourier coefficient of the non-layering mode, respectively. 
Here, we have used a strong assumption that $A^{\rm L}(\xi,\zeta,\tau)$ and $A^{\rm NL}(\xi,\zeta,\tau)$
are scaled by a common amplitude $A(\xi,\zeta,\tau)$ in the second line of Eq. (\ref{eq:neutral2}).
Expanding $A(\xi,\zeta,\tau)$ as
\begin{equation}
A(\xi,\zeta,\tau) = \epsilon A_1(\xi,\zeta,\tau) + \epsilon^2 A_2(\xi,\zeta,\tau)
+ \epsilon^3 A_3(\xi,\zeta,\tau) + \dots~, \label{eq:amplt_expan}
\end{equation}
and carrying out the weakly nonlinear analysis for the hybrid solution $\hat{\phi}_\mathrm{h}$,
we found the rescaled amplitude $\bar{A}_1(\xi,\zeta,\tau) \equiv \epsilon^{1/2} A_1(\xi,\zeta,\tau)$
for the supercritical bifurcation satisfies
\begin{equation}
\partial_\tau \bar{A}_1 = \sigma_c \bar{A}_1 + \bar{D}_1(\tau) \partial_\xi^2 \bar{A}_1
+ \bar{D}_2(\tau) \partial_\xi \partial_\zeta \bar{A}_1
+ \bar{D} \partial_\zeta^2 \bar{A}_1 + \bar{\beta} \bar{A}_1 |\bar{A}_1|^2
\label{eq:hybrid_GL3_red}
\end{equation}
at $O(\epsilon^3)$, where $\bar{D}_1(\tau)$ and $\bar{D}_2(\tau)$ are the time dependent
diffusion coefficients. Similarly, we found the higher order equation of
$\check{A}(\xi,\zeta,\tau) \equiv \epsilon^{1/2} \{A_1(\xi,\zeta,\tau)
+\epsilon A_2(\xi,\zeta,\tau)+\epsilon^2 A_3(\xi,\zeta,\tau)\}$ as
\begin{equation}
\partial_\tau \check{A} =
\sigma_c \check{A} + \bar{D}_1(\tau) \partial_\xi^2 \check{A} +
\bar{D}_2(\tau) \partial_\xi \partial_\zeta \check{A}
+ \bar{D} \partial_\zeta^2 \check{A} + \bar{\beta} \check{A} |\check{A}|^2 +
\epsilon \bar{\gamma} \check{A} |\check{A}|^4 + O(\epsilon^3)
\label{eq:hybrid_GL5_red}
\end{equation}
for the subcritical bifurcation. The time dependent diffusion coefficients $\bar{D}_1(\tau)$
and $\bar{D}_2(\tau)$ whose explicit forms are given by Eqs. (64) and (65) in Ref. \cite{saitoh1}
decay to zero as time goes on. Therefore, Eqs. (\ref{eq:hybrid_GL3_red}) and (\ref{eq:hybrid_GL5_red})
are respectively reduced to Eqs. (\ref{eq:red_TDGL3_layer}) and (\ref{eq:GL5_red}) in the long time limit.
\section{Discrete element method (DEM) simulation}\label{sec:dem}
In this section, we perform the discrete element method (DEM) simulation for
a two-dimensional granular shear flow to compare the results with the weakly nonlinear
analysis. In Sec. \ref{sub:setup}, we introduce our setup and in Sec. \ref{sub:sb},
we show the time evolution of the density field obtained from the DEM simulation,
where the typical transient
dynamics can be reproduced. In Sec. \ref{sub:vg}, we exhibit the time evolution of
the velocity fields and the granular temperature, and in Sec. \ref{sub:cm}, we compare
the results of the DEM simulation with the numerical solution of the TDGL equation.
In the following, we use the same units of mass, length and time as those in the weakly
nonlinear analysis.

In Eq. (\ref{eq:hybrid_GL3_red}), $\bar{\beta}<0$ for $\nu_0<0.245$ where the supercritical
bifurcation is expected \cite{saitoh1}. If $0.245<\nu_0<0.275$, $\bar{\beta}>0$ and
$\bar{\gamma}<0$, thus Eq. (\ref{eq:hybrid_GL5_red}) should be used and the subcritical
bifurcation is expected. Unfortunately, $\bar{\beta}>0$ and $\bar{\gamma}>0$ for
$\nu_0>0.275$ and neither Eqs. (\ref{eq:hybrid_GL3_red}) nor (\ref{eq:hybrid_GL5_red})
can be used. Therefore, we exhibit our numerical results with $\nu_0=0.18$ and $0.26$
for the supercritical and subcritical cases, respectively.
\subsection{Setup}\label{sub:setup}
We adopt the linear spring-dashpot model, where the normal force between the colliding
two particles is given by $f_n = k_n\delta-\eta_n\dot{\delta}$ with the overlap $\delta$ and the
relative speed $\dot{\delta}$. For simplicity, we ignore the tangential contact force,
because we have already verified the results are unchanged for the realistic value of
the friction coefficient by introducing the effective restitution coefficient \cite{yj,saitoh}.
In our simulation, we adopt that the spring and viscosity constants are respectively
$k_n=500 mU^2/d^2$ and $\eta_n=1.0mU/d$. In this case, the normal restitution coefficient given by
\begin{equation}
e = \exp\left[-\frac{\pi}{\sqrt{2mk_n/\eta_n^2-1}}\right] \label{eq:e}
\end{equation}
becomes $e\simeq0.9$ whose value may not be sufficiently large to ensure elastic limit \cite{DEM1,DEM2}.
We adopt that the periodic boundary condition and the Lees-Edwards boundary condition with the relative speed $U$
for the boundaries of the $\xi$- and $\zeta$-axes, respectively. Then, we
randomly distribute $N=8192$ particles in a $L^\ast\times L^\ast$ square box with the
dimensionless system size $L^\ast\equiv L/d=189$ ($\nu_0=0.18$) and $155$ ($\nu_0=0.26$),
respectively, and randomly distribute the initial velocities around the linear velocity
profile with the dimensionless shear rate $\epsilon\simeq 10^{-2}$.
\subsection{Shear band formation}\label{sub:sb}
\begin{figure}
\includegraphics[width=18cm]{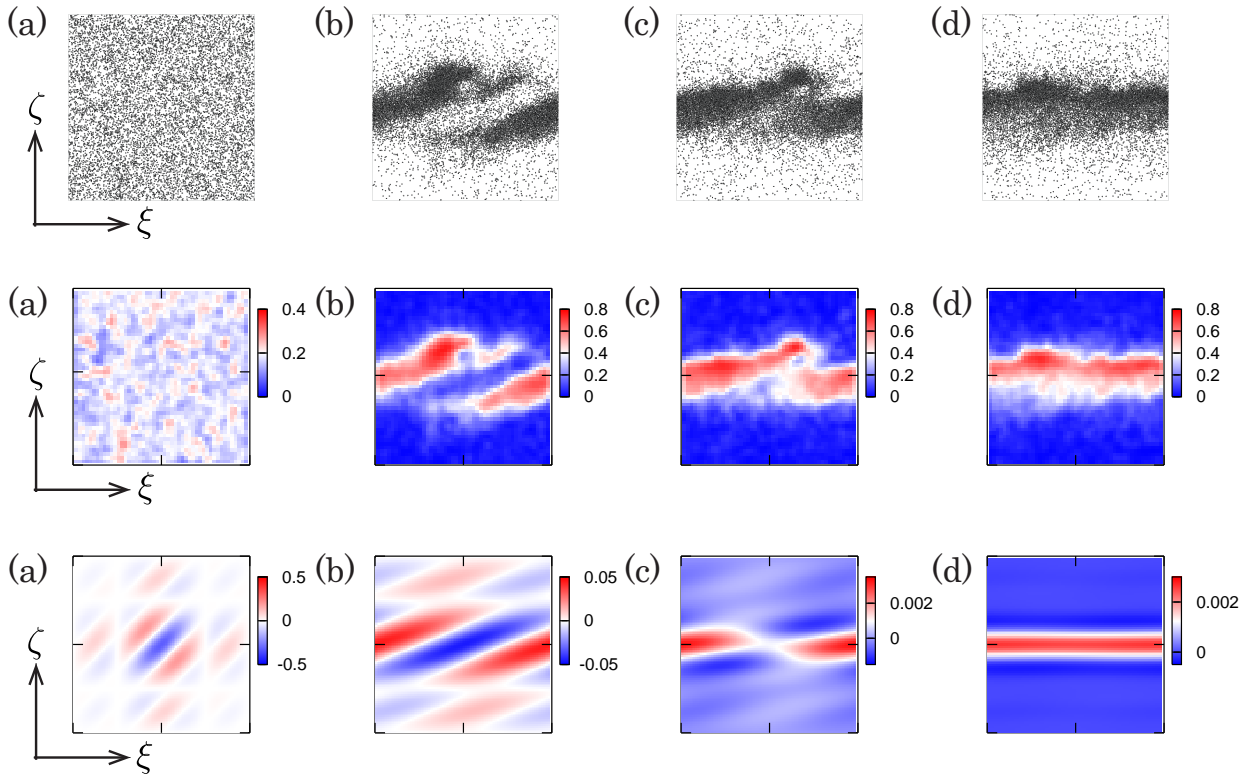}
\caption{(Color online) \emph{Upper panel}: Time evolution of particles in the DEM simulation,
where $\nu_0=0.18$. \emph{Middle panel}: Time evolution of $\nu_\mathrm{DEM}(\mathbf{z},\tau)$.
\emph{Lower panel}: Numerical solution of Eq. (\ref{eq:hybrid_GL3_red}).
Here, the dimensionless time corresponds to (a) $0$, (b) $4.8$, (c) $11.2$ and (d) $20.0$, respectively.
\label{fig:dem}}%
\end{figure}

Figure \ref{fig:dem} (\emph{upper panel}) displays the time evolution of particles
in the DEM simulation for $\nu_0=0.18$. The corresponding hydrodynamic fields can be obtained by
the coarse graining (CG) procedure developed by Goldhirsch et. al.
\cite{coarse0,coarse1,coarse2,coarse3,coarse4,coarse5,coarse6,coarse7,coarse8,coarse9},
where the CG function is defined as $\psi(\mathbf{z})=e^{-\mathbf{z}^2}/\pi$ at
$\mathbf{z}=(\xi,\zeta)$. Figure \ref{fig:dem} (\emph{middle panel}) shows the time
evolution of the area fraction defined as
\begin{equation}
\nu_\mathrm{DEM}(\mathbf{z},\tau)=\frac{\pi}{4}\sum_{i=1}^N\psi(\mathbf{z}-\mathbf{z}_i)~,\label{eq:cgdn}
\end{equation}
where $\mathbf{z}_i=(\xi_i,\zeta_i)$ is the dimensionless position of $i$-th disk.
Figure \ref{fig:dem} (\emph{lower panel}) shows the numerical solution of Eq. (\ref{eq:hybrid_GL3_red}).

In Fig. \ref{fig:dem}, a typical transient dynamics exhibits that (a) the fluctuation
with the short wave length is suppressed, (b) clusters are generated and merged, and (c)
the shear band is generated and the system reaches a steady state. Such transient dynamics
of shear band is qualitatively similar to the numerical solution of Eq. (\ref{eq:hybrid_GL3_red}).
We should stress that these results cannot be explained by neither the one-dimensional TDGL equation
nor zero-dimensional Stuart-Landau equation obtained by the ordinary weakly nonlinear analysis
\cite{shukla1,shukla2,shukla3}.
\subsection{Velocity fields and granular temperature}\label{sub:vg}
\begin{figure}
\includegraphics[width=18cm]{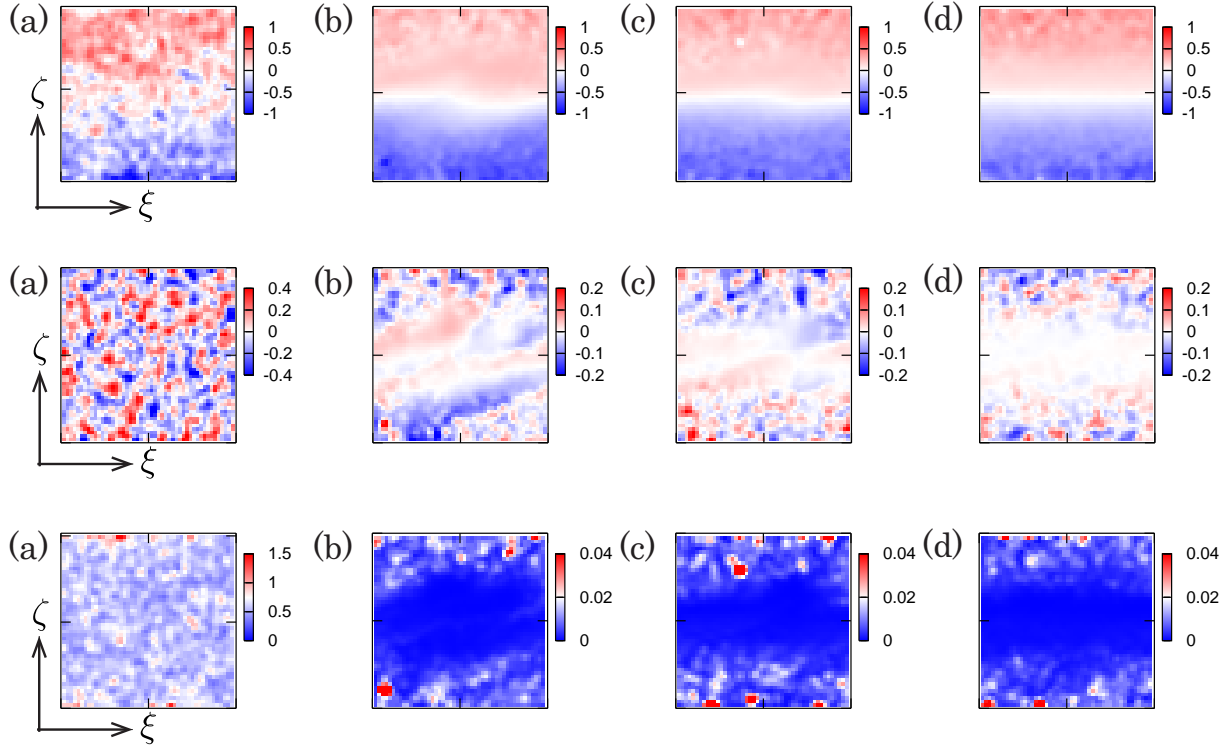}
\caption{(Color online)
\emph{Upper panel}: Time evolution of $u_\mathrm{DEM}(\mathbf{z},\tau)$.
\emph{Middle panel}: Time evolution of $w_\mathrm{DEM}(\mathbf{z},\tau)$.
\emph{Lower panel}: Time evolution of $\theta_\mathrm{DEM}(\mathbf{z},\tau)$.
Here, the dimensionless time corresponds to (a) $0$, (b) $4.8$, (c) $11.2$ and (d) $20.0$, respectively.
\label{fig:cgfd}}%
\end{figure}

The velocity fields and the granular temperature are defined as
\begin{eqnarray}
\mathbf{u}_\mathrm{DEM}(\mathbf{z},\tau)&=&\frac{\sum_i\mathbf{v}_i\psi(\mathbf{z}-\mathbf{z}_i)}
{\sum_i\psi(\mathbf{z}-\mathbf{z}_i)}~,\label{eq:cgvv}\\
\theta_\mathrm{DEM}(\mathbf{z},\tau)&=&\frac{\sum_i\mathbf{V}_i^2\psi(\mathbf{z}-\mathbf{z}_i)}
{2\sum_i\psi(\mathbf{z}-\mathbf{z}_i)}~,\label{eq:cgtm}
\end{eqnarray}
respectively, where $\mathbf{v}_i$ and $\mathbf{V}_i=\mathbf{v}_i-\mathbf{u}_\mathrm{DEM}(\mathbf{z}_i,\tau)$
are the dimensionless velocity of the $i$-th particle and the dimensionless local velocity, respectively.
Figures \ref{fig:cgfd} (\emph{upper panel}), (\emph{middle panel}) and (\emph{lower panel}) display the time
evolution of $u_\mathrm{DEM}(\mathbf{z},\tau)$, $w_\mathrm{DEM}(\mathbf{z},\tau)$ and
$\theta_\mathrm{DEM}(\mathbf{z},\tau)$, respectively, where $u_\mathrm{DEM}(\mathbf{z},\tau)$ and
$w_\mathrm{DEM}(\mathbf{z},\tau)$ are respectively the $\xi$ and $\zeta$ components of
$\mathbf{u}_\mathrm{DEM}(\mathbf{z},\tau)$. As time goes on, $u_\mathrm{DEM}(\mathbf{z},\tau)$
in the $\zeta$ direction deviates from the linear profile and $w_\mathrm{DEM}(\mathbf{z},\tau)$ is almost homogeneous.
The time evolution of $\theta_\mathrm{DEM}(\mathbf{z},\tau)$ is accompanied with
$\nu_\mathrm{DEM}(\mathbf{z},\tau)$, where $\theta_\mathrm{DEM}(\mathbf{z},\tau)$
is lower in the dense region and higher in the dilute region.
\subsection{Comparison of the TDGL equation with the DEM simulation}\label{sub:cm}
To test the quantitative validity of the TDGL equation, we compare the numerical solution
with the results of DEM simulation. 
At first, we average out $\nu_\mathrm{DEM}(\mathbf{z},\tau)$,
$u_\mathrm{DEM}(\mathbf{z},\tau)$, $w_\mathrm{DEM}(\mathbf{z},\tau)$ and
$\theta_\mathrm{DEM}(\mathbf{z},\tau)$ over the $\xi$ direction and take sample averages
from the different $100$ time steps. 
Then, the hydrodynamic fields are written as one-dimensional forms $\nu_\mathrm{DEM}(\zeta,\tau)$, $u_\mathrm{DEM}(\zeta,\tau)$,
$w_\mathrm{DEM}(\zeta,\tau)$ and $\theta_\mathrm{DEM}(\zeta,\tau)$, respectively.
Because $\nu_\mathrm{DEM}(\zeta,\tau)$ and $\theta_\mathrm{DEM}(\zeta,\tau)$ are
approximately symmetric at $\zeta=0$, we introduce
\begin{eqnarray}
\bar{\nu}_\mathrm{DEM}(\zeta,\tau)&\equiv&\frac{1}{2}\left\{\nu_\mathrm{DEM}(\zeta,\tau)
+\nu_\mathrm{DEM}(-\zeta,\tau)\right\}\hspace{0.5cm}(0<\zeta<L^\ast/2)~,\label{eq:barnu}\\
\bar{\theta}_\mathrm{DEM}(\zeta,\tau)&\equiv&\frac{1}{2}\left\{\theta_\mathrm{DEM}(\zeta,\tau)
+\theta_\mathrm{DEM}(-\zeta,\tau)\right\}\hspace{0.5cm}(0<\zeta<L^\ast/2)~,\label{eq:barth}
\end{eqnarray}
respectively. On the other hand, the velocity fields are approximately antisymmetric
at $\zeta=0$ and we also introduce
\begin{eqnarray}
\bar{u}_\mathrm{DEM}(\zeta,\tau)&\equiv & \frac{1}{2}\left\{u_\mathrm{DEM}(\zeta,\tau)
-u_\mathrm{DEM}(-\zeta,\tau)\right\}\hspace{0.5cm}(0<\zeta<L^\ast/2)~,\label{eq:barux}\\
\bar{w}_\mathrm{DEM}(\zeta,\tau)&\equiv & \frac{1}{2}\left\{w_\mathrm{DEM}(\zeta,\tau)
-w_\mathrm{DEM}(-\zeta,\tau)\right\}\hspace{0.5cm}(0<\zeta<L^\ast/2)~,\label{eq:baruy}
\end{eqnarray}
respectively.

In the weakly nonlinear analysis, the hydrodynamic fields are given by the summation
of the base state $\phi_0=(\nu_0,\zeta,0,\theta_0)$ and the hybrid solution $\hat{\phi}_\mathrm{h}$.
At first, we project $\hat{\phi}_\mathrm{h}$ on the $\zeta$-axis as
\begin{equation}
\hat{\phi}_\mathrm{h}(\zeta,\tau)\simeq\bar{\bar A}(\zeta,\tau)\phi^{\rm L}_{q_c}
e^{iq_\zeta(\tau)\zeta}+\mathrm{c.c.}~, \label{eq:Abarbar}
\end{equation}
where $q_\zeta(\tau)\equiv q_c-\tau$ is the $\zeta$ component of $\mathbf{q}(\tau)$
\cite{qzeta} and we ignore $\phi^{\rm NL}_{\mathbf{q}(\tau)}$, because $\phi^{\rm NL}_{\mathbf{q}(\tau)}$
exponentially decays to zero and the following results are unchanged even if we take into
account $\phi^{\rm NL}_{\mathbf{q}(\tau)}$. We note that $\phi^{\rm L}_{q_c}$ is defined as
$\phi^{\rm L}_{q_c}=\left(\nu_{q_c},iu_{q_c},iw_{q_c},\theta_{q_c}\right)^\mathrm{T}$
with the imaginary unit $i$, where $\nu_{q_c}$, $u_{q_c}$, $w_{q_c}$
and $\theta_{q_c}$ are the Fourier coefficients of the area fraction, the velocity fields
$u$ and $w$, and the granular temperature, respectively, and they
are given in our previous paper \cite{saitoh1}. If we ignore the higher order terms
in Eq. (\ref{eq:amplt_expan}), $\bar{\bar A}(\zeta,\tau)$ may be given by the numerical
solution of Eq. (\ref{eq:hybrid_GL3_red}) projected on the $\zeta$-axis.
Then, the hydrodynamic fields are given by
$\phi_\mathrm{TDGL}(\zeta,\tau)=\phi_0+\hat{\phi}_\mathrm{h}(\zeta,\tau)$,
where each component of $\phi_\mathrm{TDGL}(\zeta,\tau)$ is written as
\begin{eqnarray}
\nu_\mathrm{TDGL}(\zeta,\tau) &=& \nu_0+2\nu_{q_c}\bar{\bar A}(\zeta,\tau)\cos(q_\zeta(\tau)\zeta)~,\label{eq:tdgl_nu}\\
u_\mathrm{TDGL}(\zeta,\tau) &=& \zeta-2u_{q_c}\bar{\bar A}(\zeta,\tau)\sin(q_\zeta(\tau)\zeta)~,\label{eq:tdgl_ux}\\
w_\mathrm{TDGL}(\zeta,\tau) &=& -2w_{q_c}\bar{\bar A}(\zeta,\tau)\sin(q_\zeta(\tau)\zeta)~,\label{eq:tdgl_uy}\\
\theta_\mathrm{TDGL}(\zeta,\tau) &=& \theta_0+2\theta_{q_c}\bar{\bar A}(\zeta,\tau)\cos(q_\zeta(\tau)\zeta)~,
\label{eq:tdgl_th}
\end{eqnarray}
respectively, where the factor $2$ comes from the complex conjugate.

\begin{figure}
\includegraphics[width=16cm]{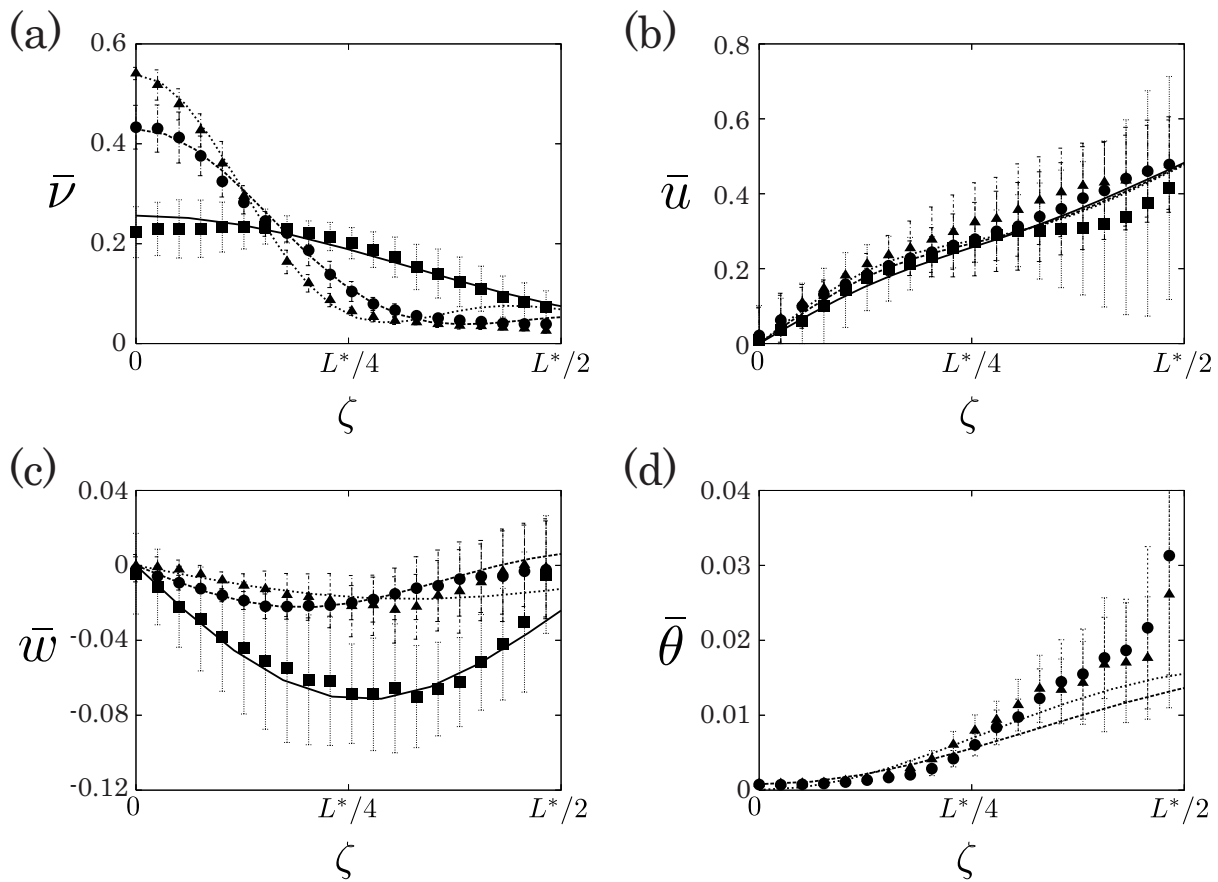}%
\caption{Time evolution of
(a) $\bar{\nu}_\mathrm{DEM}(\zeta,\tau)$ and $\bar{\nu}_\mathrm{TDGL}(\zeta,\tau)$,
(b) $\bar{u}_\mathrm{DEM}(\zeta,\tau)$ and $\bar{u}_\mathrm{TDGL}(\zeta,\tau)$,
(c) $\bar{w}_\mathrm{DEM}(\zeta,\tau)$ and $\bar{w}_\mathrm{TDGL}(\zeta,\tau)$,
(d) $\bar{\theta}_\mathrm{DEM}(\zeta,\tau)$ and $\bar{\theta}_\mathrm{TDGL}(\zeta,\tau)$,
respectively, for the supercritical case ($\nu_0=0.18$), where the solid squares and the solid
lines correspond to the dimensionless time $4.8$, the solid circles and the hashed
lines correspond to the dimensionless time $11.2$, and the solid triangles and
the dotted lines correspond to the dimensionless time $20.0$, respectively.
Here, we have introduced the scaling factors $\tau^\ast\simeq 0.14$, $a^\ast_\nu\simeq 0.24$,
$a^\ast_u\simeq 0.02$,
$a^\ast_w\simeq 1.96$ and $a^\ast_\theta\simeq 0.02$, respectively, and we also use
$\zeta_\nu^\ast(\tau)\simeq 1.6, 0.9, 0.75$, $\zeta_u^\ast(\tau)\simeq 0.8, 0.8, 0.8$,
$\zeta_w^\ast(\tau)\simeq 1.5, 1.1, 1.8$ at the the dimensionless time $4.8$, $11.2$
and $20.0$, respectively, and $\zeta_\theta^\ast(\tau)\simeq 1.6, 1.35$ at the
dimensionless time $11.2$ and $20.0$, respectively. It should be noted that we
do not show the result of the granular temperature at the dimensionless time $4.8$,
because it homogeneously distributed around $\theta_0$ and the fluctuation is too
large to plot in the same figure.
Here, the relative standard deviations defined as Eq. (\ref{eq:err})
are (a) $0.09$, (b) $0.07$, (c) $0.10$ and (d) $0.35$, respectively.
\label{fig:sup}}%
\end{figure}

\begin{figure}
\includegraphics[width=16cm]{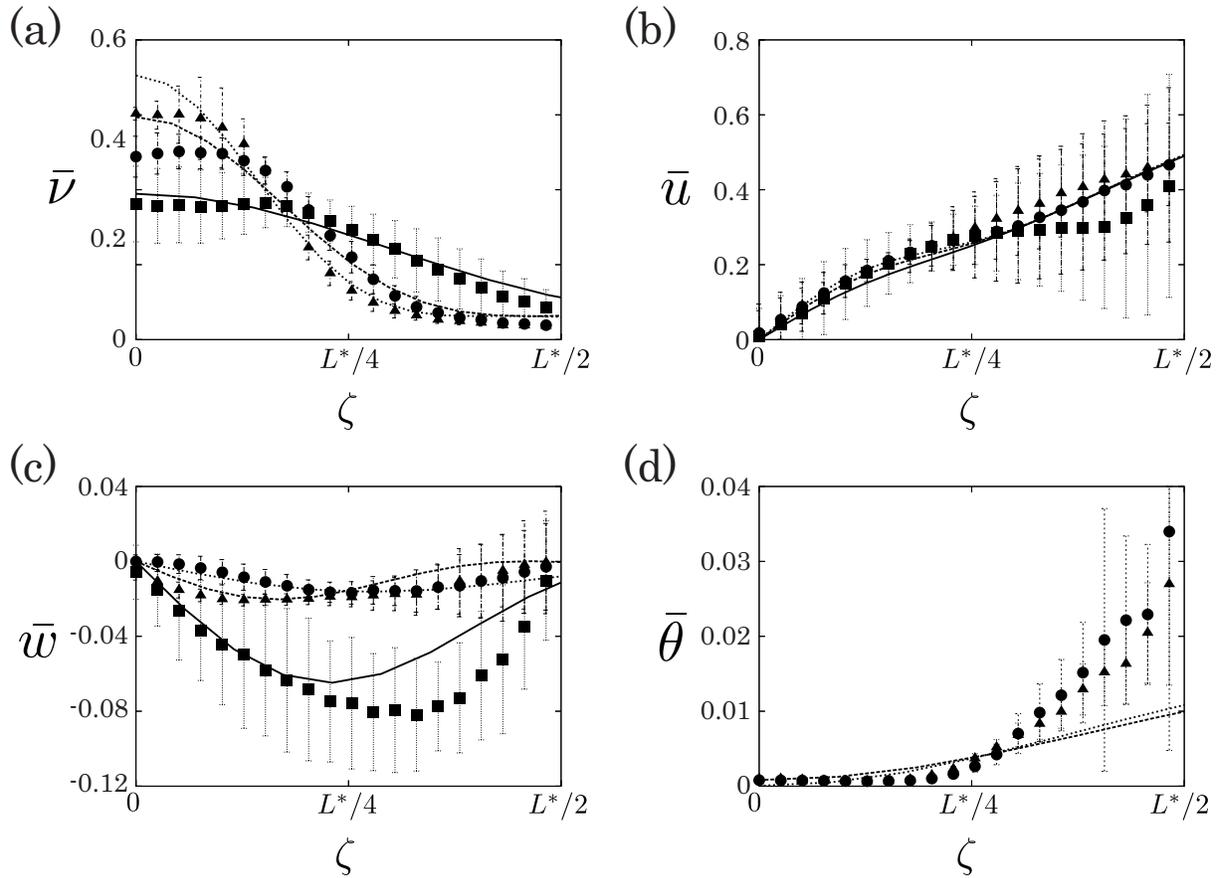}%
\caption{Time evolution of
(a) $\bar{\nu}_\mathrm{DEM}(\zeta,\tau)$ and $\bar{\nu}_\mathrm{TDGL}(\zeta,\tau)$,
(b) $\bar{u}_\mathrm{DEM}(\zeta,\tau)$ and $\bar{u}_\mathrm{TDGL}(\zeta,\tau)$,
(c) $\bar{w}_\mathrm{DEM}(\zeta,\tau)$ and $\bar{w}_\mathrm{TDGL}(\zeta,\tau)$,
(d) $\bar{\theta}_\mathrm{DEM}(\zeta,\tau)$ and $\bar{\theta}_\mathrm{TDGL}(\zeta,\tau)$,
respectively, for the subcritical case ($\nu_0=0.26$), where we have used the same
dimensionless time and the same scaling factors $\tau^\ast$, $a^\ast_\nu$, $a^\ast_u$,
$a^\ast_w$, $a^\ast_\theta$, $\zeta_\nu^\ast(\tau)$, $\zeta_u^\ast(\tau)$, $\zeta_w^\ast(\tau)$
and $\zeta_\theta^\ast(\tau)$ in Fig. \ref{fig:sup}.
Here, the relative standard deviations defined as Eq. (\ref{eq:err})
are (a) $0.10$, (b) $0.07$, (c) $0.40$ and (d) $0.43$, respectively.
\label{fig:sub}}%
\end{figure}

Figures \ref{fig:sup} and \ref{fig:sub} display the time evolution of the hydrodynamic
fields for the supercritical case ($\nu_0=0.18$) and the subcritical case ($\nu_0=0.26$),
respectively, where the symbols represent Eqs. (\ref{eq:barnu})-(\ref{eq:baruy}) obtained
by the DEM simulation and the lines represent the scaling functions
\begin{equation}
\bar{X}_\mathrm{TDGL}(\zeta,\tau)\equiv
a^\ast_XX_\mathrm{TDGL}(\zeta/\zeta^\ast_X(\tau),\tau/\tau^\ast)
\hspace{0.5cm}(X=\nu,u,w,\theta) \label{eq:scaling}
\end{equation}
with the scaling factors $a^\ast_X$, $\zeta^\ast_X(\tau)$ and $\tau^\ast$, respectively.
We quantify the difference between Eqs. (\ref{eq:barnu})-(\ref{eq:baruy}) and Eq. (\ref{eq:scaling})
by introducing the relative standard deviation
\begin{equation}
\mathrm{Err.} \equiv \sqrt{\frac{\left(\bar{X}_\mathrm{DEM}-\bar{X}_\mathrm{TDGL}\right)^2}
{\bar{X}_\mathrm{TDGL}^2}} \hspace{0.5cm}(X=\nu,u,w,\theta)~, \label{eq:err}
\end{equation}
where we omit the arguments $(\zeta,\tau)$.
In Fig. \ref{fig:sup}(a)-(c), $\bar{\nu}_\mathrm{TDGL}(\zeta,\tau)$, $\bar{u}_\mathrm{TDGL}(\zeta,\tau)$
and $\bar{w}_\mathrm{TDGL}(\zeta,\tau)$ are quantitatively agreed with $\bar{\nu}_\mathrm{DEM}(\zeta,\tau)$,
$\bar{u}_\mathrm{DEM}(\zeta,\tau)$ and $\bar{w}_\mathrm{DEM}(\zeta,\tau)$, respectively,
where $\mathrm{Err.}$ is less than or equal to $0.1$.
In Fig. \ref{fig:sub} (a) and (b), $\bar{\nu}_\mathrm{TDGL}(\zeta,\tau)$ and $\bar{u}_\mathrm{TDGL}(\zeta,\tau)$
are quantitatively agreed with $\bar{\nu}_\mathrm{DEM}(\zeta,\tau)$ and $\bar{u}_\mathrm{DEM}(\zeta,\tau)$,
respectively.
We should note that we could not get any reasonable agreements between the $\zeta$
component of the velocity field even in a numerical solution of a set
of the granular hydrodynamic equations and the result of DEM simulation in our
previous work \cite{saitoh}.
We can also see the qualitative agreements in the $\zeta$ component of the velocity
field for the subcritical case (Fig. \ref{fig:sub}(c)) and the granular temperature
for the supercritical and subcritical cases (Figs. \ref{fig:sup}(d) and \ref{fig:sub}(d)),
where $\mathrm{Err.}$ is less than or equal to $0.43$.
\section{Discussion and conclusion}\label{sec:summ}
In this paper, we examine the validity of the TDGL equation for a two-dimensional
sheared granular flow from the comparison with the results of the DEM simulation
by the CG method. The results of the TDGL equation, at least, qualitatively agree
with the results of the DEM simulation. Such transient dynamics cannot be reproduced
by neither the one dimensional TDGL equation nor the zero dimensional Stuart-Landau
equation derived by the ordinary weakly nonlinear analysis. We also obtain that the
velocity fields and the granular temperature qualitatively agree with the solution
of the TDGL equation.

We compare the one dimensional hydrodynamic fields obtained from the DEM simulation
with the scaled forms of the numerical solution of the TDGL equation, where
we find the quantitative agreements in the area fraction and the $\xi$ component
of the velocity field. In the supercritical regime, we also find the quantitative
agreement in the $\zeta$ component of the velocity field. We can also observe the
qualitative agreements in the $\zeta$ component of the velocity field for the
subcritical case and the granular temperature for both the supercritical and subcritical
cases.
In our previous work, the hydrodynamic fields obtained from the DEM simulation are
reasonably explained by the numerical solutions of the granular hydrodynamic equations
by Jenkins and Richmann except for $w(\mathbf{z},\tau)$~\cite{jr1,jr2,saitoh}.
In the present work, even though we need to introduce the scaling factors, the results
of the DEM simulation is qualitatively reproduced by the numerical solution of the TDGL
equation. It is needless to say that more precise analyses will be important to remove
the scaling factors. In addition, quantitative comparison with the DEM simulations in quasi elastic
limit should be done in our future studies.

In conclusion, the numerical solution of the TDGL equation can qualitatively explain
the time evolution of the hydrodynamic fields obtained by the DEM simulation.
\begin{acknowledgments}
This work was financially supported by an NWO-STW VICI grant.
Numerical computation in this work was carried out at the Yukawa Institute Computer Facility.
\end{acknowledgments}
%

\begin{thebibliography}{82}
\bibitem[{qze()}]{qzeta}
\bibinfo{note}{Here,
  $\mathbf{q}(\tau)=\mathbf{q}_c+\delta\mathbf{q}(\tau)\equiv (\delta
  q,q_c-\epsilon t\delta q)$ and we use $\delta q\sim\epsilon$, thus the
  $\zeta$ component of $\mathbf{q}(\tau)$ is given by
  $q_\zeta(\tau)=q_c-\tau$.}
\end{thebibliography}

\end{document}